\documentclass[%
reprint,
 amsmath,amssymb,
 aps,
 prl,
]{revtex4-1}

\usepackage{graphicx}% Include figure files
\usepackage{dcolumn}% Align table columns on decimal point
\usepackage{bm}% bold math
\usepackage{hyperref}% add hypertext capabilities
\usepackage{color}

\begin{document}

\preprint{APS/123-QED}

\title{Strain Control and Layer-Resolved Switching of Negative Capacitance in BaTiO$_3$/SrTiO$_3$ Superlattices}
% Force line breaks with \\

\author{Raymond Walter$^{1,2,*}$, Charles Paillard$^{1,3}$, Sergey Prosandeev$^{1,4}$ and L. Bellaiche$^{1}$}%
% \email{rwalter@email.uark.edu}
 \affiliation{$^{1}$ Department of Physics and Institute for Nanoscience and Engineering, University of Arkansas, Fayetteville, Arkansas 72701, USA \\
 $^{2}$ Department of Mathematical Sciences, University of Arkansas, Fayetteville, Arkansas 72701, USA \\
 $^{3}$ Laboratoire Structures, Propri\'et\'es et  Mod\'elisation des Solides, CentraleSup\'elec, UMR CNRS 8580, Universit\'e Paris-Saclay, 91190 Gif-sur-Yvette, France \\
$^{4}$ Institute of Physics and Physics Department of Southern Federal University, Rostov-na-Donu 344090, Russia} 

\date{\today}% It is always \today, today,

\begin{abstract}
Negative capacitance in BaTiO$_3$/SrTiO$_3$ superlattices is investigated by Monte Carlo simulations in an atomistic effective Hamiltonian model, using fluctuation formulas for responses to the local macroscopic field that incorporates depolarizing fields. 
We show epitaxial strain can tune the negative capacitance of the BaTiO$_3$ ferroelectric layer and the overall capacitance of the system over a broad temperature range. 
In addition, we predict and explain an original switching of the negative capacitance from the BaTiO$_3$ layer to the SrTiO$_3$ layer at low temperatures for intermediate strains.
Our results indicate how the diffusive character of the multidomain transition in these superlattices improves their viability for capacitance applications.
\end{abstract}

\maketitle

\section{\label{sec:intro}I. Introduction}
Negative capacitance (NC), although thermodynamically unstable, was recently realized in ferroelectrics, either transiently upon switching the ferroelectric polarization~\cite{Khan2015,Hoffmann2016, Hoffmann2018} or statically in SrTiO$_3$/PbTiO$_3$ superlattices~\cite{Zubko2016} and ferroelectric nanodots~\cite{Lukyanchuk2019}. Realizing NC in a ferroelectric embedded between semiconductors is one of the simplest, most promising routes available to defeat the Boltzmann tyranny that plagues transistor energetic consumption, and to enable the design of more efficient transistors~\cite{Salahuddin2008}. 

Two distinct origins have been proposed for static or low-frequency NC. The first mechanism uses a monodomain ferroelectric in series with a stiff dielectric to force the ferroelectric into a paraelectric state~\cite{Iniguez2019} with negative curvature of the Landau potential~\cite{Krowne2011}. The second relies on the domain pattern and incomplete screening (through the appearance of a dielectric dead layer or finite-length metallic screening)~\cite{Bratkovsky2001} of the polarization induced by domain wall motion~\cite{Iniguez2019,Lukyanchuk2018}. 

The aforementioned works have shown the main mechanism causing NC relies on the generation of a larger depolarizing field response in the ferroelectric than the overall applied electric field. Most works so far focused on the relative thickness of the ferroelectric and dielectric layers~\cite{Zubko2016} or on electrostatic screening~\cite{Bratkovsky2001,Ponomareva2007a,Ponomareva2007b} to induce NC. In this work, we propose epitaxial strain as an alternative handle to control NC. We demonstrate how strain in (BaTiO$_3$)$_m$/(SrTiO$_3$)$_n$ superlattices  (BTO/STO SLs) can tune the magnitude of NC and its temperature range. 
Varying Ba and Sr compositions in titanate perovskites has proven to be effective to obtain large, tunable dielectric permittivities for capacitors, either in bulk solid solution~\cite{Walizer2006} or films~\cite{Dawber2005,Okatan2009,Cole2003,Lisenkov2007}.
We show that \textit{(i)} strain and the different resulting dipolar configurations tune NC in (BaTiO$_3$)$_8$/(SrTiO$_3$)$_2$ superlattices, and, \textit{(ii)} under large compressive strain and low temperature, a transfer of NC from the BTO to the STO layer occurs that, to the best of our knowledge, has never been reported. We also interpret NC in terms of responses to the local macroscopic electric field that incorporates depolarizing fields. 
\section{\label{sec:method}II. Method and System}
Our work uses the effective Hamiltonian model of Ref.~\cite{Walizer2006} that expresses the total energy as $H_{eff} = H_{ave} \left( \left\lbrace {u}_i \right\rbrace, \left\lbrace v_i \right\rbrace, \eta_H \right) + H_{loc} \left( \left\lbrace u_i \right\rbrace, \left\lbrace v_i \right\rbrace, \left\lbrace \sigma_i \right\rbrace \right)$ in terms of a few degrees of freedom: the local soft mode in a unit cell $i$, ${u}_i$, proportional to the polarization; inhomogeneous strain describing the deformation of unit cell $i$, $v_i $; and the homogeneous strain $\eta_H$ of the supercell. $H_{ave}$ represents the average total energy of a virtual $\left\langle \text{Ba}_{0.5} \text{Sr}_{0.5} \right\rangle \text{TiO}_3$ crystal, and $H_{loc}$ represents the energetic perturbation due to the chemical distribution of Ba and Sr cations ($\sigma_i = +1$ for Ba and $\sigma_i = -1$ for Sr in cell $i$).
This model accurately described Curie temperatures and phase diagrams in disordered and ordered (Ba, Sr)TiO$_3$ systems~\cite{Walizer2006,Lisenkov2007,Choudhury2011}, with the proviso that it treats SrTiO$_3$ as (Ba, Sr)TiO$_3$ with a small Ba concentration of 15\% by predicting an unstrained bulk paraelectric-to-ferroelectric transition around 100~K.

 We solve this model using Metropolis Monte Carlo (MC) simulations in a $12 \times 12 \times 10$ supercell to mimic a (BaTiO$_3$)$_8$/(SrTiO$_3$)$_2$ superlattice grown along the pseudo-cubic [001] direction.
 The supercell is field cooled under a $200$~kV.cm$^{-1}$ from $1000$~K to $25$~K by $25$~K steps using $2\times 10^5$ MC sweeps, and then to $5$~K in $5$~K steps using $10^6$ sweeps. 
 The field is then removed, and the system is heated from 5~K to 25~K (5~K steps), and from 25~K to 1525~K (25~K steps) using $10^6$ sweeps. 
Thermodynamic averages are taken over the last $8\times 10^5$ sweeps; static dielectric susceptibilities are estimated based on linear response theory, using correlators as described below.

\begin{figure}[t]
	\centering
 	\includegraphics[scale=0.96]{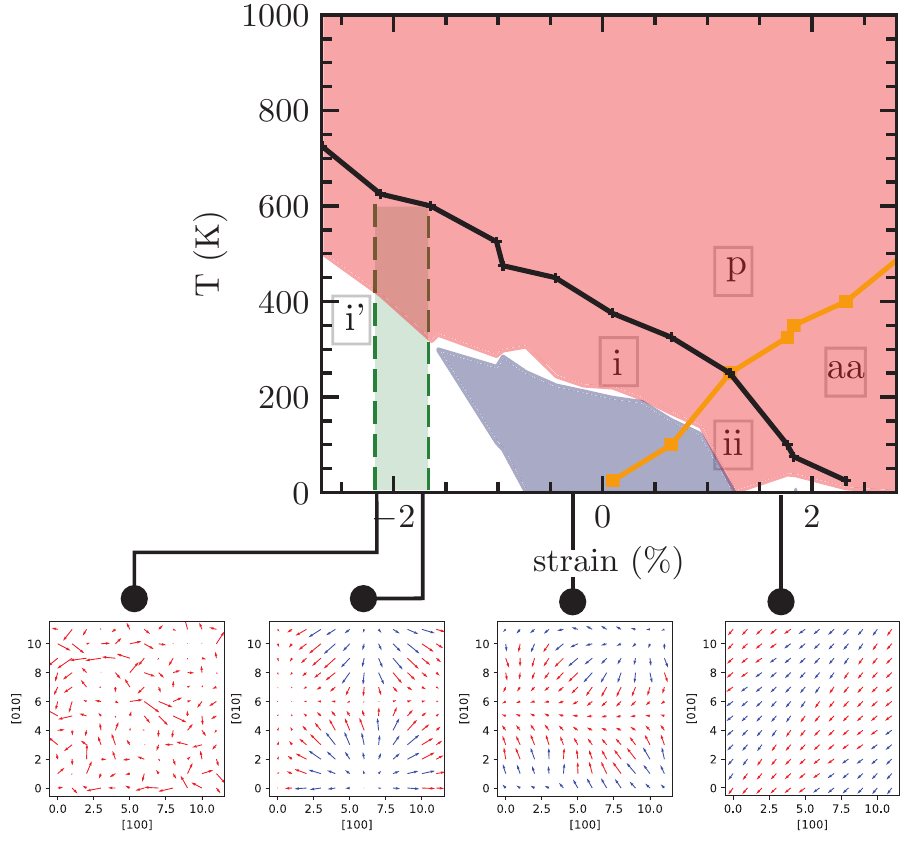}
	\caption{Phase diagram of the (BaTiO$_3$)$_8$/(SrTiO$_3$)$_2$ superlattice, with multidomain out-of-plane polar phase $i$, monodomain out-of-plane polar phase $i^{\prime}$, orthorhombic polar phase with in-plane polarization $aa$, phase $ii$ having both in-plane polarization and out-of-plane polar domains, and high-temperature paraelectric phase $p$. The black and yellow lines separate these phases. The green shaded area demarcates the stability region of polar nanobubbles. The red and blue shaded areas represent regions of NC for the BTO and STO layers. In-plane dipolar patterns are depicted for 1.77, $-0.45$, $-1.56$ and $-2.02\%$ epitaxial strains and shows up and down dipoles colored in red and blue respectively.}
	\label{fig:PhaseDiagram}
\end{figure}

We observe different phases depending on strain and temperature in the superlattice, as depicted in Figure \ref{fig:PhaseDiagram}. 
At high temperature, a paraelectric phase $p$ with disordered dipole patterns is the most stable state. 
At low temperature and moderate and larger tensile strain ($>1.65$\% around 300~K), the superlattice is orthorhombic with in-plane polarization ($aa$ phase in Figure \ref{fig:PhaseDiagram}); between large tensile and small compressive strain ($-0.35$ to 2.40\% at 5~K) and low temperature, alternating out-of-plane polar domains develop in the $ii$ phase in addition to an in-plane polarization. 
At moderate and larger compressive strain, no in-plane polar order exists, but out-of-plane alternating polar domains remain in the $i$ phase. 
At large compressive strain, a monodomain phase $i^{\prime}$ with out-of-plane polarization becomes stable and is separated from the $i$ phase by a polar nanobubble phase (green area and snapshot in Figure \ref{fig:PhaseDiagram}). 
The phase diagram agrees qualitatively with previous reports~\cite{Lisenkov2007,Choudhury2011}, except that monodomain phase $i^{\prime}$ and polar nanobubbles are presently found, likely because our studied system has smaller overall Sr concentration than in those studies. 
Note that polar nanobubbles, despite their prediction more than ten years ago~\cite{Lai2006}, only recently have been observed experimentally in ferroelectric superlattices~\cite{Zhang2017}.
\section{\label{sec:theory}III. Theory}
We focus on the dielectric response of the superlattice in different regions of the phase diagram.
Care must be taken when defining the different layer-resolved dielectric susceptibilities and dielectric permittivities. 
Indeed, unlike the global dielectric permittivity of a solid modeled with periodic boundary conditions (\textit{i.e.}, neglecting surface effects), the layers of that periodic system each experience a depolarizing field in response to an applied electric field. 
Then the macroscopic electric field in the layer (which is not the microscopic dipole field in our effective Hamiltonian~\cite{Ponomareva2005}) divides into a local and an applied part: $E_{\text{layer}}=E_{\text{loc}}+E_{\text{ext}}$.
One can thus define \textit{external} and \textit{internal} susceptibilities for a layer, 
$\chi_{layer}^{ext} = \frac{\Delta P_{\text{layer}}}{\varepsilon_0\Delta E_{\text{ext}}}$ and $\chi_{layer}^{int} = \frac{\Delta P_{\text{layer}}}{\varepsilon_0\Delta E_{\text{layer}}}$, where 
$\Delta E_{\text{layer}}$
 and $\Delta P_{\text{layer}}$  are the respective changes of total electric field and polarization in the layer in reaction to an applied external electric field $\Delta E_{\text{ext}}$~\cite{Ponomareva2007a,Ponomareva2007b}. 
All polarizations and electric fields are considered in the stacking direction of the superlattice.
We calculate these responses according to the linear response formulas:
  \begin{subequations}
    \begin{align}
\chi _ {layer} ^ { ext } 
&=\frac { 1 } { \varepsilon_0 }\frac { V } { k _ { B } T } \left[ \left\langle  P _ { \mathrm{layer} } P _ {\mathrm{tot}}\right\rangle - \left\langle P _ { \mathrm{layer}} \right\rangle \left\langle P _ {\mathrm{tot}}  \right\rangle\right] \label{local external}\\
\chi _ {layer } ^ { int } 
&=\chi _ {layer} ^ {ext}\Bigg/\frac { \partial \left\langle E _ {\mathrm{layer}} \right\rangle }{ \partial E _ {\text{ext}}} \label{local internal}.
    \end{align}
  \end{subequations}
The total volume factor $V$ for the system prevents finite-size scaling issues when local and global responses are compared; $T$ and $k_B$ refer to temperature and Boltzmann's constant; angular brackets indicate thermodynamic averages. 
One can define an internal dielectric permittivity $\varepsilon^{int}_{layer}$ for a layer in addition to the total dielectric permittivity $\varepsilon_{tot}$ of the system,\begin{subequations}
    \begin{align}
\varepsilon^{int}_{layer} & =  1 + \chi_{layer}^{int},\label{epsilon internal}\\
 \varepsilon_{tot}             & =  1 +\frac { 1 } { \varepsilon_0 }\frac { V } { k _ { B } T } \left[ \left\langle  P _ { \mathrm{tot} } ^2\right\rangle - \left\langle P _ { \mathrm{tot}} \right\rangle^2\right]
   \end{align}
  \end{subequations}
\noindent %For the internal dielectric response, w
We rely upon a separate fluctuation formula for the response of the internal electric field to the external electric field: 
  \begin{subequations}
    \begin{align}
      \frac{\partial\left\langle E _ { \mathrm{layer}  } \right\rangle }{\partial E _ {\text{ext}}} & =  1+ \frac { V } { k _ { B } T } [ \left\langle E _ { \mathrm{loc} } P _ {\mathrm{tot}} \right\rangle - \left\langle E _ { \mathrm{loc}} \right\rangle\left\langle P _ { \mathrm{tot} } \right\rangle ]  \label{field response} 
     \\ & =  \varepsilon_{tot} - \chi_{layer}^{ext} \quad \mathrm{(periodic}\hspace{0.2em}\mathrm{superlattice)}. \label{field response SL} 
    \end{align}
  \end{subequations}
\noindent Equation (\ref{field response}) applies under general electrical boundary conditions that are not necessarily periodic. 
In a periodic superlattice as considered here, $E _ { \mathrm{loc}  }=(P _ { \mathrm{tot} }-P _ { \mathrm{layer} })/\varepsilon_{0}$~\cite{Zubko2016}; then, upon substituting Equation  (\ref{field response SL}) into Equation  (\ref{local internal}), we recover the formula for the internal relative dielectric permittivity $\varepsilon_{layer}^{int}=\varepsilon_{tot}/(\varepsilon_{tot} - \chi_{layer}^{ext})$ in Ref.~\cite{Zubko2016}. 
Equations (\ref{local internal}), (\ref{epsilon internal}), and (\ref{field response}) imply one can achieve NC if $\chi_{layer}^{ext}+\frac{\partial\left\langle E _ { \mathrm{layer}  } \right\rangle }{\partial E _ {\text{ext}}}<0$, when the change in residual depolarizing field in the layer $\Delta E _ { \mathrm{loc} }$ caused by an applied field $E _ { \mathrm{ext} }$ is larger in magnitude than $E _ { \mathrm{ext} }+\Delta P _ { \mathrm{layer} }/\varepsilon_{0}$~\cite{Ponomareva2007a,Ponomareva2007b}. 
%In the Supplemental Material, we derive these formulas, and 
%compare direct and statistical approaches to calculating $\varepsilon^{int}_{layer}$. 
%We also derive a version of the sum rule for capacitances in series, which lets us precisely interpret NC in terms of a negative $\varepsilon^{int}_{layer}$.
\subsection{A. Derivations}
We derive the above formulas and a version of the sum rule for (inverse) capacitances in series that lets us precisely interpret negative capacitance in terms of a negative internal dielectric permittivity $\varepsilon^{int}_{layer}$. In this subsection only, we will assume a superlattice with $M$ layers and thicknesses $l_i, .. i=1,\dots,M,$ and overall thickness $L$. We adopt the same notation as in Ref.~\cite{Zubko2016}. All polarizations and electric fields are considered in the stacking direction of the superlattice.

The macroscopic electric field in the layer divides into a local and an applied part: $E_{\text{layer}}=E_{\text{loc}}+E_{\text{ext}}$. However, we now label the layer by index $i$, \emph{e.g.}, $E_{\text{i}}=E_{\text{i,loc}}+E_{\text{ext}}$
The \textit{external} and \textit{internal} susceptibilities for a layer are
$\chi_{i}^{ext} = \frac{\Delta P_{\text{i}}}{\varepsilon_0\Delta E_{\text{ext}}}$ and $\chi_{i}^{int} = \frac{\Delta P_{\text{i}}}{\varepsilon_0\Delta E_{\text{i}}}$, where 
$\Delta E_{\text{i}}$
 and $\Delta P_{\text{i}}$  are respectively the change of total electric field and polarization in the layer $i$ in reaction to an applied external electric field $\Delta E_{\text{ext}}$~\cite{Ponomareva2007a,Ponomareva2007b}. 

We prove the various fluctuation formulas for local polarization.
%  \begin{subequations}
%    \begin{align}
%\chi _ {i} ^ { ext } 
%&=%\frac { 1 } { \varepsilon_0 }\frac { \partial \left\langle P _ { \mathrm{layer}} \right\rangle } { \partial E _ {\text{ext}} } =
%\frac { 1 } { \varepsilon_0 }\frac { V } { k _ { B } T } \left[ \left\langle  P _ {i} P _ {\mathrm{tot}}\right\rangle - \left\langle P _ {i} \right\rangle \left\langle P _ {\mathrm{tot}}  \right\rangle\right] \label{local external}\\
%\chi _ {i} ^ { int } 
%&=%\frac { \partial \left\langle P _ { \mathrm{layer} } \right\rangle } { \partial  \left\langle E _ { \mathrm{layer} } \right\rangle  } =
%\chi _ {i} ^ {ext}\Bigg/\frac { \partial \left\langle E _ {i} \right\rangle }{ \partial E _ {\text{ext}}} \label{local internal}\\
%      \frac{\partial\left\langle E _ { i  } \right\rangle }{\partial E _ {\text{ext}}} & =  1+ \frac { V } { k _ { B } T } [ \left\langle E _ { \mathrm{i,loc} } P _ {\mathrm{tot}} \right\rangle - \left\langle E _ { \mathrm{i,loc}} \right\rangle\left\langle P _ { \mathrm{tot} } \right\rangle ]  \label{field response}\\
%      \varepsilon^{int}_{i} & =  1 + \chi_{i}^{int}, 
%       \end{align}
%  \end{subequations}
%where for periodic superlattices~\cite{Zubko2016}:
%\begin{equation}
%E _ { \mathrm{i,loc}  }=(P _ { \mathrm{tot} }-P _ {i})/\varepsilon_{0}.
%\label{localField}
%\end{equation}
Our effective Hamiltonian can be written in the form 
$$
H = H ^ { ( 0) } - V P _ {\mathrm{tot}} E _ {\mathrm{ext}},
$$
\noindent where $V P _ {\mathrm{tot}}$ is the dipole of the finite sample, $E _ {\mathrm{ext}}$ is the applied field, and $H^{(0)}$ is the energy functional of the system in the absence of an applied field. To extend to the case of nanostructures subject to lower dimensional electrical boundary conditions, we would also separate out the maximal depolarizing field contributions~\cite{Ponomareva2007a}.
%From the definition of average polarization, we know that $L ^ { - 1} \sum _ { i = 1} ^ { M } l _ { i } P _ { i } = P _ {tot}$ holds, where $P _ { i }$ labels the average polarization in layer $i$. Our effective Hamiltonian can be rewritten as 
%$$
%H = H ^ { ( 0) } - V \frac{1}{L} \Big(l _ { i }  P_ { i }+\sum _ { i \neq j} l _ { i }P _ { i } \Big) \cdot E _ {\mathrm{ext}}.
%$$ 
We use linear response theory~\cite{frenkel2001understanding}. We consider the perturbation of an observable $A$:
\begin{equation*}
\langle A\rangle_{0}%+\langle\Delta A\rangle
=\frac{\int \mathrm{d} \Gamma \exp \left[-\left(H ^ { ( 0) }- V P _ {\mathrm{tot}}E _ {\mathrm{ext}}\right)/(k _ { B } T)\right] A}{\int \mathrm{d} \Gamma \exp \left[-\left(H ^ { ( 0) }- V P _ {\mathrm{tot}}E _ {\mathrm{ext}}\right)/(k _ { B } T)\right]},
\end{equation*}
where $\Gamma$ indicates the phase space coordinates.
In the limit of small fields, we have the derivative 
\begin{eqnarray*}
\left(\frac{\partial\langle A\rangle}{\partial E _ {\mathrm{ext}}}\right)_{E _ {\mathrm{ext}}=0}&=&\frac { 1 } { k _ { B } T } \left\{\langle A V P _ {\mathrm{tot}}\rangle-\langle A\rangle\langle V P _ {\mathrm{tot}}\rangle\right\}\\
&=&\frac { V } { k _ { B } T }\left\{\langle A P _ {\mathrm{tot}}\rangle-\langle A\rangle\langle P _ {\mathrm{tot}}\rangle\right\}.
\end{eqnarray*}

If layer polarization $P_i$ is the observable, then
  \begin{equation}
\chi _ {i} ^ { ext }= \frac { 1 } { \varepsilon_0 } \frac { \partial \left\langle P _ {  i  } \right\rangle }{ \partial E _ { \mathrm{ext} } }= \frac { 1 } { \varepsilon_0 }\frac { V } { k _ { B } T } \left[ \left\langle  P _ { i } P _ {\mathrm{tot}}\right\rangle - \left\langle P _ {i} \right\rangle \left\langle P _ {\mathrm{tot}}  \right\rangle\right]. \label{local external 2}
  \end{equation}
  
\noindent The layer internal dielectric response formula is obtained by the chain rule~\cite{Ponomareva2007a}:
\begin{equation}
\chi _ { i } ^ { ext } = \frac { 1 } { \varepsilon_0 }\frac { \partial \left\langle P _ {  i  } \right\rangle } { \partial \left\langle E _ {  i  } \right\rangle } \frac { \partial \left\langle E _ {  i  } \right\rangle }{ \partial E _ { \mathrm{ext} } }= \chi _ { i } ^ { int }  \frac { \partial \left\langle E _ {  i  } \right\rangle }{ \partial E _ { \mathrm{ext} } }.
\end{equation}
To evaluate the response of the macroscopic electric field of the layer to the externally applied field, one starts by observing that the macroscopic field has two parts, one of which is the external field, so we have
  \begin{equation}
      \frac{\partial\left\langle E _ {i} \right\rangle }{\partial E _ {\text{ext}}} =  1+ \frac{\partial\left\langle E _ { \mathrm{i,loc}  } \right\rangle }{\partial E _ {\text{ext}}}.
  \end{equation}
Then we need a cumulant formula for the local contribution only. This amounts to choosing $E _ { \mathrm{i,loc}  }$ as the observable in the above linear response formulas. 
One can obtain Equation  (\ref{field response}).
%In a periodic system like ours, $\sum _ { i = 1} ^ { M }  E _ { \mathrm{i,loc}  }=0$, we can rely on the familiar trick of applying a fictitious field that couples to the perturbation whose contribution to the Hamiltonian vanishes, then using linear response theory to calculate it as a response function~\cite{}. 
To extend to general electrical boundary conditions, the depolarizing field contributions~\cite{Ponomareva2007a,Ponomareva2005} can be added to the local field.

Interestingly, besides the described above cumulant approach to the calculation of the local internal response, one can also use a direct approach. Indeed, starting from the definitions given in Ref.~\cite{Ponomareva2007a} and, by employing $E _ { \mathrm{loc}  }=(P _ { \mathrm{tot} }-P _ { \mathrm{layer} })/\varepsilon_{0}$~\cite{Zubko2016}, one can recover the direct approach formula~\cite{Zubko2016} for the calculation of the internal dielectric permittivity in a periodic superlattice:
\begin{eqnarray}\label{local internal epsilon}
  \epsilon_{i}^ { \text{int}  }&=&1+\chi _ { i } ^ { \text{int} }=1+\frac{1}{\epsilon_0} \frac { \Delta P _ { i } } { \Delta E _ i } \nonumber\\
  &=& 1+\frac{\Delta P _ { i }}{\left( \Delta P _ {\mathrm{tot}} - \Delta P _ { i } \right)+\epsilon_0 E _ { \text{ext} }} \nonumber\\
  &=& \frac{\Delta P _ {\mathrm{tot}} +\epsilon_0 E _ { \text{ext} }}{\left( \Delta P _ {\mathrm{tot}} - \Delta P _ { i } \right)+\epsilon_0 E _ { \text{ext} }} \nonumber\\
  &=& \Big(\chi _ { tot } ^ { \text{ext} }+1\Big)/\Big(\Big(\chi _ { tot } ^ { \text{ext} }-\chi _ { i} ^ { \text{ext} }\Big)+1\Big)\nonumber\\
  &=& { \varepsilon_{\mathrm{tot}}  }/\Big({ \varepsilon_{\mathrm{tot}}  }-\chi_{i}^ { \text{ext}  }\Big),
\end{eqnarray}
\noindent A cumulant approach like ours was, in fact, hinted at in Ref.~\cite{Zubko2016}, but was not elaborated upon or used; we find the cumulant approach more valuable.

Equation (\ref{local internal epsilon}) makes it easy to verify the series capacitance rule. Note discrete differences make it easy to verify the series capacitance rule:
\begin{eqnarray*}
\sum_{i=1}^{M}l_{i }\Big(\varepsilon _ { tot } ^ { \text{ext} }-\chi _ { i} ^ { \text{ext} }\Big)  E _ { \text{ext} }&=&\sum _ { i = 1} ^ { M } l _ { i } (1+\frac { \Delta P _ {\mathrm{tot}}- \Delta P _ { i } } {  E _ { \text{ext} }})E _ { \text{ext} }\nonumber\\
 &=& L\Big(E _ { \text{ext} }+\Delta P _ {\mathrm{tot}}-\frac{1}{L}\sum _ { i = 1} ^ { M } l _ { i }  \Delta P _ { i } \Big)\nonumber\\
 &=&L\Big(E _ { \text{ext} }+\Delta P _ {\mathrm{tot}}-\Delta P _ {\mathrm{tot}}\Big)\nonumber\\
 &=&LE _ { \text{ext} }.
\end{eqnarray*}
Then for fixed cross-section $A$ perpendicular to the stacking direction,
\begin{eqnarray}
%\sum _ { i = 1} ^ { M } \frac{1}{C_i}=
\sum _ { i = 1} ^ { M } \frac{l _ { i }}{A\epsilon_0} \frac{1}{\epsilon_{i}^ { \text{int}  }}&=&\sum _ { i = 1} ^ { M } \frac{l _ { i }}{A\epsilon_0} \frac{\Big(\varepsilon_{\mathrm{tot}}-\chi_{i}^ { \text{ext}  }\Big)} { \varepsilon_{\mathrm{tot}}  }\nonumber\\&=&\frac{L}{A\varepsilon_0} \frac{1}{\epsilon_{\mathrm{tot}}}=\frac{1}{C_{tot}}.\label{series rule}
\end{eqnarray}
To be clear, we are proving this ``internal" series capacitance sum rule, not merely assuming it.
We can say a few things about this sum rule. Satisfying this formal sum suggests that we can reasonably interpret a layer with internal dielectric permittivity $\epsilon_{i}^ { \text{int}  }$ to contribute a capacitance of $C_i=\frac{A\epsilon_0}{l _ { i }}\epsilon_{i}^ { \text{int}  }$. 
Also, we can view a ferroelectric-dielectric superlattice system differently from the traditional picture of a capacitor filled with several layers of dielectric material with known external dielectric permittivities. Rather, each layer of material has an internal dielectric permittivity that is determined by the local electrical environment in the adjacent layers \cite{Ponomareva2007a,Zubko2016,Bratkovsky2001}.
%(Not necessarily simultaneously determined, since dynamical slowing of charge carriers is an important mechanism for negative capacitance~\cite{Khan2015}.) 
The local internal dielectric permittivities together determine the global external dielectric permittivity.
%
%Third, the suggestion that electrical behavior in each layer are affected by those in all other layers resembles the decomposition of Born effective charges into contributions from each layer in \cite{Wu2006}. Our notions can be rigorously treated in a Wannier function framework as in the rest of the modern theory of electrical polarization.
\section{IV. Negative Capacitance Optimization and Switching}
\begin{figure*}
	\centering
 	\includegraphics[scale=0.85]{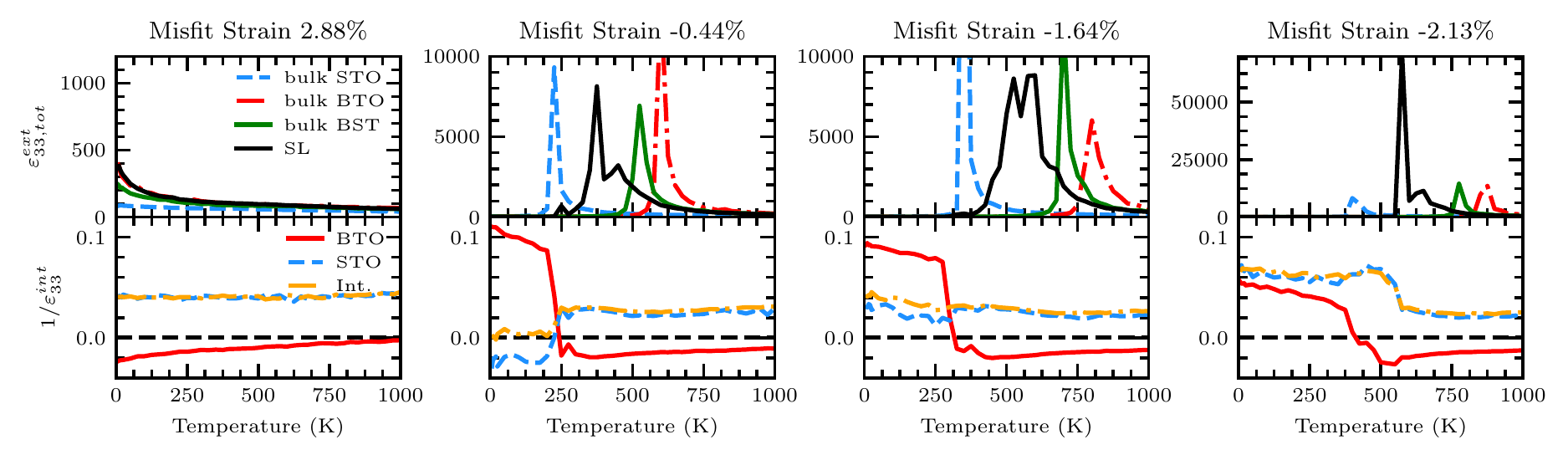}
	\caption{(Upper panels) total external dielectric permittivity of the superlattice (black line), dielectric permittivity of bulk STO (blue dashed), bulk BTO (red dashed dotted), and bulk BST, 80\% Ba (green solid); (lower panels) inverse internal dielectric permittivity for BTO (red), STO (blue dashed) and Interfacial (orange dashed dotted) layers within the BTO/STO superlattice.}
	\label{fig:Epsilons}
\end{figure*}

The upper panels of Figure \ref{fig:Epsilons} report the external out-of-plane dielectric permittivities of strained bulk BaTiO$_3$ (BTO), SrTiO$_3$ (STO) and disordered (Ba$_{0.8}$Sr$_{0.2}$)TiO$_3$ (BST) in red, blue, and green respectively; we compare with the external dielectric response of our BTO/STO superlattice (black solid line) for different strains. 
The blue, red, and green lines in those upper panels refer to the dielectric permittivities of \textit{separate bulk} calculations, not to the external dielectric permittivities of the slabs, which equal that of the overall superlattice~\cite{Zubko2016}. 
The lower panels of Figure \ref{fig:Epsilons} report the inverse internal dielectric response, $1/\varepsilon_{33}^{int}$, in the BTO, STO, and Interfacial layers using red, blue, and orange lines, respectively. For technical reasons in our model, mentioned below, there are two Interfacial layers, one STO layer, and seven BTO layers; we consider all layers of the same type together.
Starting with a very large 2.88\% tensile strain, the superlattice only experiences a $p$ to $aa$ transition, according to Figure \ref{fig:PhaseDiagram}. 
We thus do not expect to observe a peak in the \textit{out-of-plane} external dielectric permittivity $\varepsilon_{33,tot}^{ext}$, as confirmed in the rightmost panel in Figure \ref{fig:Epsilons}. 
Looking at $1/\varepsilon_{33}^{int}$, we observe NC in the BTO layers through the whole temperature range represented, while STO and Interfacial layers have a nearly constant \textit{positive} $1/\varepsilon_{33}^{int}$. 
The sum rule for
capacitances in series implies that maximizing the overall capacitance amounts to maximizing the magnitude of $1/\varepsilon_{33}^{int}$ of the ferroelectric layer \textit{at constant value} of the internal dielectric permittivity of the dielectric layers, explaining why the overall dielectric permittivity of the BTO/STO superlattice is maximum at low temperature for this tensile case, where $\left| 1/\varepsilon_{33, BTO}^{int} \right|$ is largest.

\begin{figure*}
%	\centering
	\includegraphics[trim=10 0 0 0,scale=0.98]{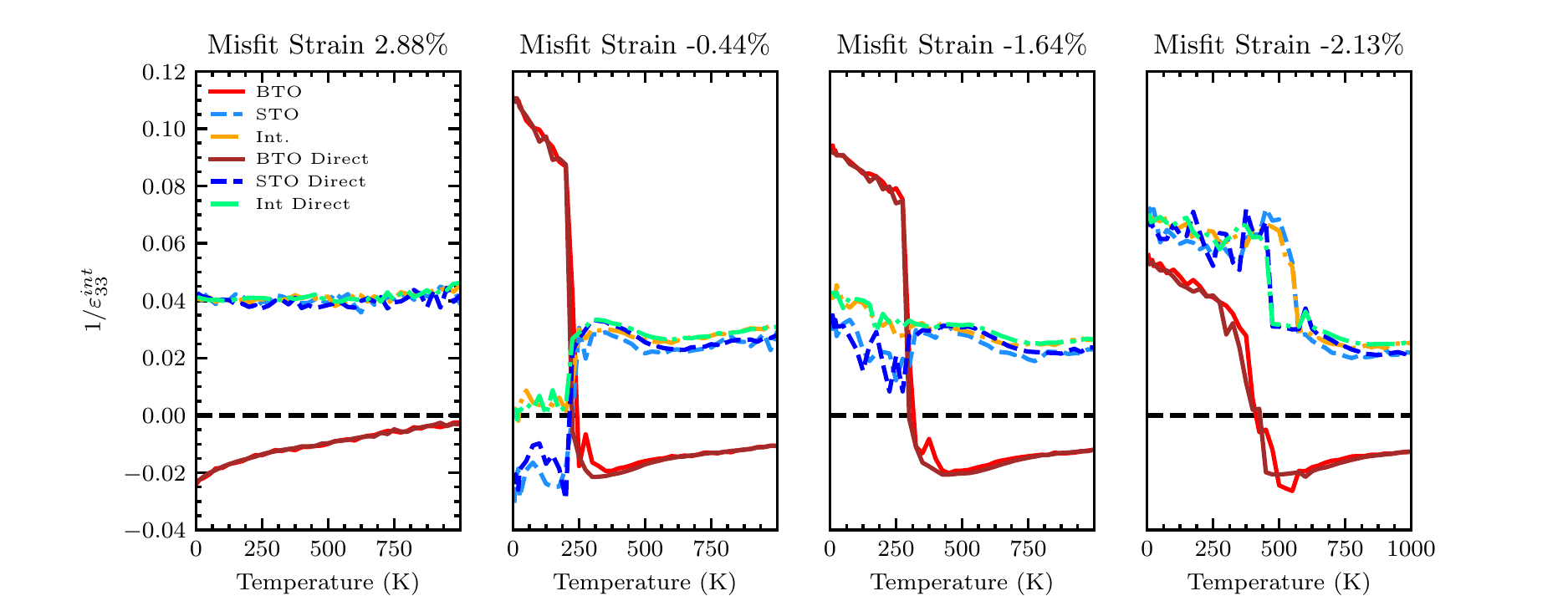}
	\caption{Inverse internal dielectric permittivity for BTO (red), STO (sky blue dashed) and interfacial (orange dashed dotted) layers calculated by cumulant formulas and for BTO (brown), STO (dark blue dashed), and Interfacial (green dashed dotted) by a direct approach within the BTO/STO superlattice.}
	\label{fig:EpsDirectVSTemp}
\end{figure*}

Note a direct approach as used in Ref.~\cite{Zubko2016} that calculates
 $\varepsilon^{int}_{layer}$ by finite differences upon applying a small electric field:
  \begin{equation}
  \varepsilon_{layer}^{int} = 1+\frac{\Delta P_{\text{layer}}}{(\Delta P_{\text{tot}}-\Delta P_{\text{layer}})+\varepsilon_0\Delta E_{\text{ext}}},\label{direct epsilon}
  \end{equation}
is in excellent agreement with the statistical approach we use; see Figure \ref{fig:EpsDirectVSTemp}.

We next turn our attention to the $-0.44\%$ strained superlattice in Figure \ref{fig:Epsilons}. According to Figure \ref{fig:PhaseDiagram}, the out-of-plane polarization must develop in an alternating multidomain structure, and, correspondingly, a broad (between 175~K and 500~K) peak is observed in the external dielectric permittivity; $\varepsilon_{33,tot}^{ext}$ is enhanced in this region relative to bulk STO. However, at temperatures lower than 175~K, \textit{i.e.}, below the peak in the external dielectric permittivity, an original inversion occurs: the internal dielectric permittivity of the ferroelectric multidomain BTO layer jumps to positive values while the STO layer now exhibits \textit{negative} dielectric permittivity. In the $i$ phase where the switching is realized, the STO layer adopts a similar multidomain pattern as the BTO layer. 

\begin{figure}[h]
%	\centering
 	\includegraphics[trim=20 0 0 0,scale=0.95]{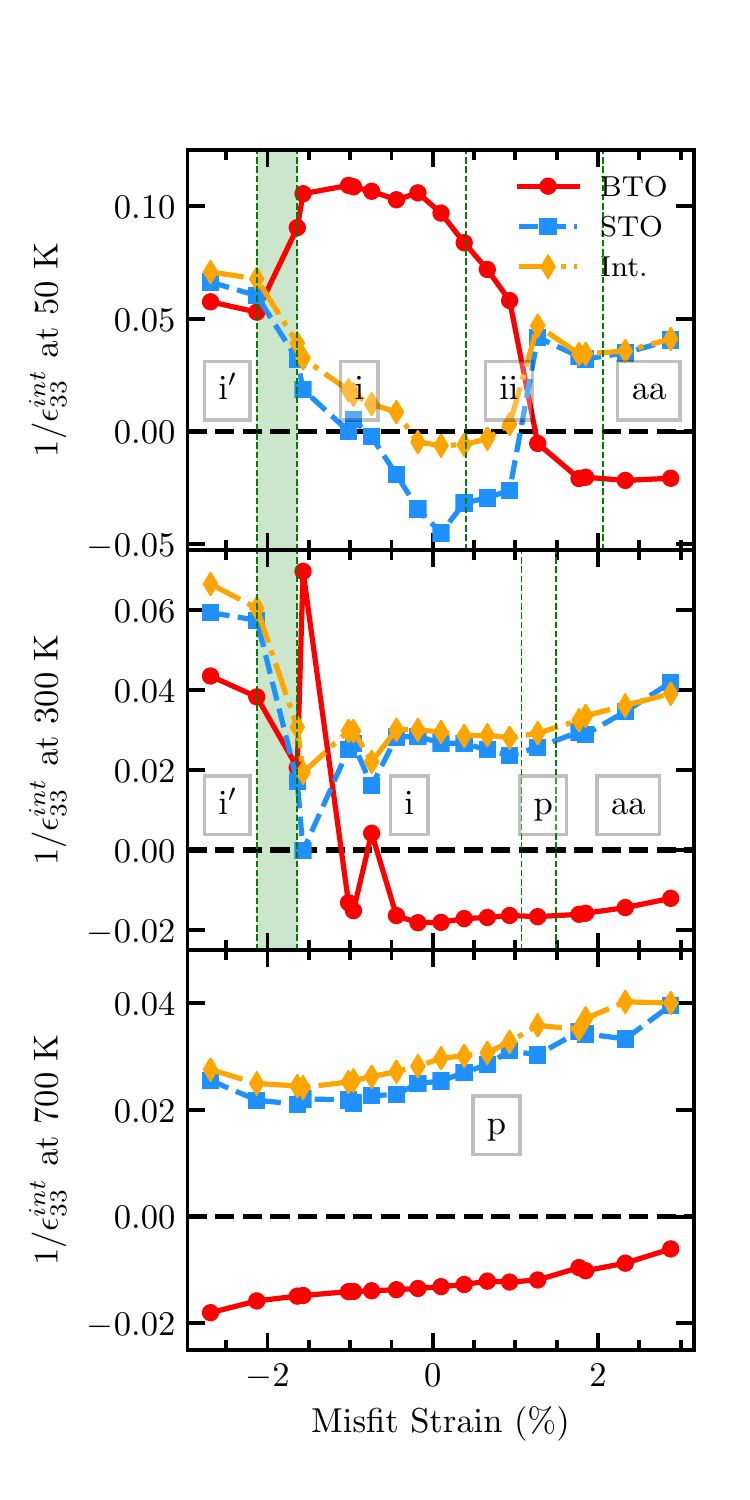}
	\caption{\label{fig:EpsVSStrain} Inverse internal dielectric permittivity of the BTO (red circles), STO (blue squares) and Interfacial (orange diamonds) layers for different strains, at 50~K (upper panel), 300~K (middle panel) and 700~K (lower panel) in the BTO/STO superlattice.}
\end{figure}

To understand this switching of the negative capacitance between the BTO and STO layers, we plot the evolution of $1/\varepsilon_{33}^{int}$ of the BTO (red circles), STO (blue squares) and Interfacial (orange diamonds) layers with strain at 50, 300 and 700~K in Figure \ref{fig:EpsVSStrain}. The STO layer shows NC only at low temperature for strains within regions of the phase diagram with an out-of-plane multidomain configuration (see blue shaded area in Figure \ref{fig:PhaseDiagram}). The upper panel of Figure \ref{fig:EpsVSStrain} also shows that switching of NC between the STO and BTO layers can be realized inside the same phase, state $i$, by changing strain. As Equation  (\ref{field response SL}) indicates, negative internal dielectric permittivities 
occur when one of the layers is much more polarizable than the overall structure~\cite{Zubko2016}. We deduce that, below the switching temperature, the dipolar fluctuations in the STO layer are more important than those in the BTO layer; there are a few ways this can happen. 
First, as shown in Figure \ref{fig:Epsilons}, bulk STO becomes ferroelectric under compressive strain \cite{Pertsev2000}; our model overcompensates slightly by predicting bulk STO exhibits out-of-plane polarization, 
since it treats unstrained bulk STO as effectively BST with 15\% Ba concentration~\cite{Walizer2006}. Then the STO layer can experience NC in the usual way for ferroelectrics, especially as the BTO layer suffers an energy penalty for developing a large polarization or depolarizing field~\cite{Zubko2016}. Recent work also relies on compressive strain to induce out-of-plane polarization for NC~\cite{Lukyanchuk2019,Cheng2019}, though not for a switching like ours.
Second, diffuse phase transitions in these superlattices imply the STO layers can exhibit long tails of large dielectric response, even in the absence of a ferroelectric transition in bulk STO, that exceed the diminished response of bulk BTO reflected by smaller responses in the BTO layers. 

\subsection{A. Local Field Responses}
Negative capacitance switching is associated with the separation of local fields and their responses in the different layers. We show the stark separation of local fields and their response in the different layers associated with the switching between the BaTiO$_3$ and SrTiO$_3$ layers of the (BaTiO$_3$)$_{8}$/(SrTiO$_3$)$_{2}$ superlattice. 
%We also outline a phenomenological view of the negative capacitance switching between the BTO and STO layers. 

\begin{figure*}
	\centering
	\includegraphics[scale=0.9]{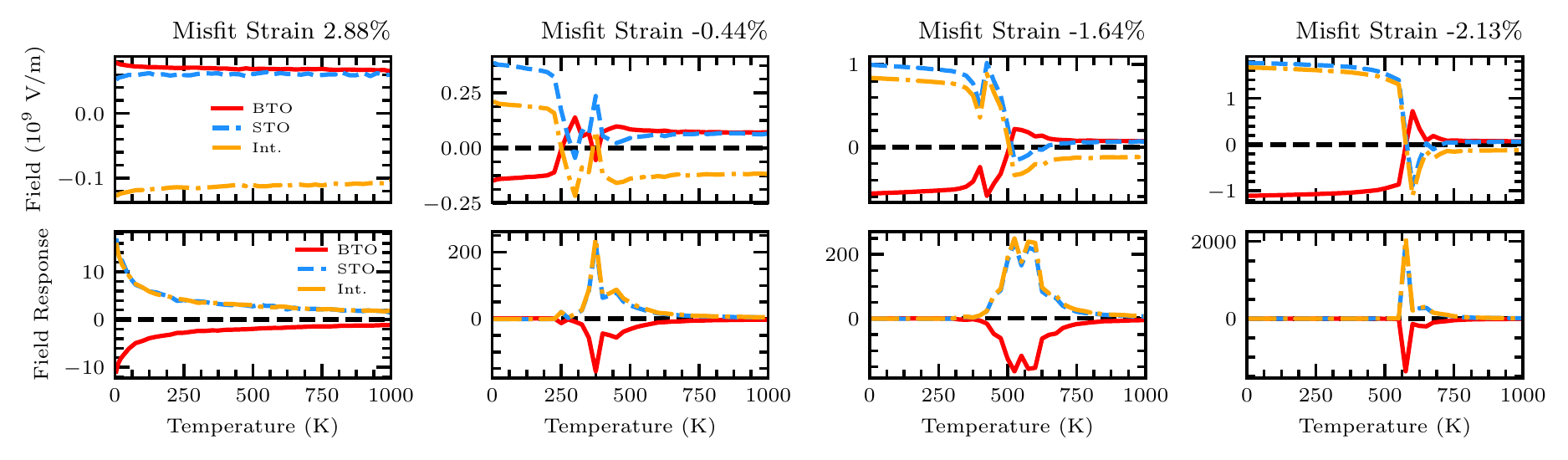}
	\caption{(Upper panels) Local contribution to macroscopic electric field experienced in the layer for BTO (red), STO (blue dashed) and interfacial (orange dashed dotted) layers within the BTO/STO superlattice; (lower panels) response of the macroscopic field experienced in the layer to an externally applied electric field for BTO (red), STO (blue dashed) and interfacial (orange dashed dotted) layers within the BTO/STO superlattice.}
	\label{fig:FieldVSTemp}
\end{figure*}

The upper panels of Figure \ref{fig:FieldVSTemp} report the local contribution to the macroscopic electric field experienced in the layer in the BTO (red), STO (blue dotted), and  Interfacial (orange dotted dashed) layers of the superlattice for several misfit strains. The lower panels of Figure \ref{fig:FieldVSTemp} report the response of these local field contributions to the externally applied field, which are calculated using correlators as in Equation  (\ref{field response}). Negative capacitance occurs in a  layer when the field response goes negative. There are a few notable features.
\begin{itemize}
\item   At higher temperatures (at all temperatures, in the case of the high tensile regime, \emph{e.g.}, 2.88\% strain), the Interfacial layers have a local field oppositely directed to and larger in magnitude than the BTO and STO layers, which are almost equal to each other.
\item However, the field responses of the BTO and STO layers are opposite in sign and, therefore, have opposite internal dielectric permittivities. Then the direction of the field does not determine the occurrence of negative capacitance. Comparison of the STO and Interfacial layers proves a similar point for the magnitude of the field and sign of the internal dielectric permittivity.
%\item Though the local fields in the STO and Interfacial layers are oppositely directed at higher temperatures, their responses are nearly equal. Then neither the direction or magnitude of the local field determines the sign of the field response and, therefore, neither determines the sign of the internal dielectric permittivity (the occurrence of negative capacitance). 
\item Outside the high tensile regime, at low temperatures, the fields in the BTO and STO layers adopt opposite orientations, generally with a larger magnitude in the STO layer.
\item At lower temperature, the field settles to an almost constant value and its response stiffens, reducing significantly in magnitude in each layer. In the BTO layer, except in the high tensile regime, this reduction in magnitude involves a switch of sign. In the $-0.44\%$ case and similarly intermediate strains, the STO and Interfacial layers exhibit a switch of sign as well.
\end{itemize}

\begin{figure*}
	\centering
	\includegraphics[scale=0.9]{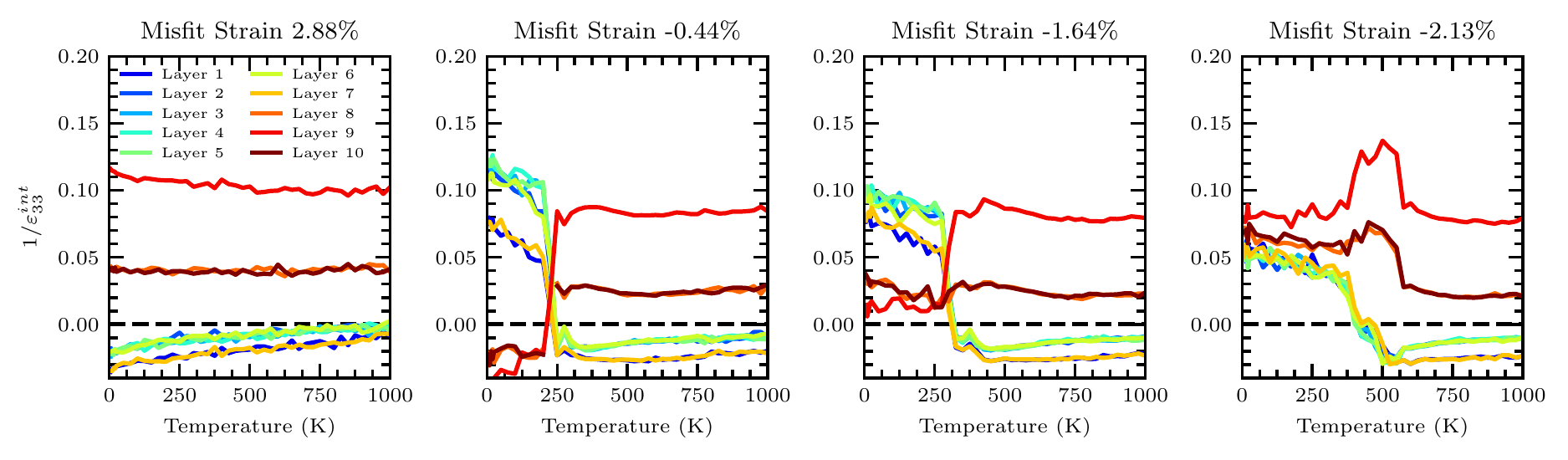}
	\caption{Inverse internal dielectric permittivity for the individual layers of the BTO/STO superlattice, where Layer 9 is the STO layer, Layers 8 and 10 are the Interfacial Layers, and the rest are BTO layers.}
	\label{fig:EpsVSTempLayer}
\end{figure*}

In this work, the two Interfacial layers have been treated together because our effective Hamiltonian treats them the same way with respect to composition-dependent epitaxial strain. (For technical details, see Ref.~\cite{Walizer2006}.) However, one of these layers (since we use Ti-centered local modes the layer means a layer of TiO$_2$) has Ba above and Sr below while the other layer has the opposite orientation, \emph{i.e.}, the composition gradient is opposite. More complicated behavior can result than in the BTO or STO layers. Therefore, it is worth examining the (inverse) internal dielectric permittivities in all ten individual layers of the superlattice in Figure \ref{fig:EpsVSTempLayer}. We are reassured to find that the two Interfacial layers exhibit the same internal dielectric permittivity. We also find that, in our case, the internal dielectric permittivities all BTO layers agree in sign throughout their temperature range, but that, when negative, the layers nearest to the interface exhibit the largest inverse values and hence deliver the greatest amplification for overall capacitance.

\subsection{B. Nanobubbles, Monodomains, And Enhancement Over Bulk}
The inversion of the NC between the BTO and STO layers occurs for a range of compressive strains that extends to the nanobubble phase boundary in Figure \ref{fig:PhaseDiagram}. 
Within the nanobubble phase (see $-1.64\%$ compressive strain), it appears that, despite a significant decrease of $1/\varepsilon_{33,STO}^{int}$ at 260~K, the STO layer never experiences NC when the BTO layer switches from negative to positive capacitance near 260~K. 
The concurrent decrease in magnitude of $1/\varepsilon_{33,STO}^{int}$ and $1/\varepsilon_{33,BTO}^{int}$ prevents the global dielectric permittivity $\varepsilon_{33,tot}$ in Figure \ref{fig:Epsilons} to peak when BTO switches from negative to positive internal dielectric permittivity. 
That is, both internal dielectric permittivities (and associated capacitances) of STO and BTO diverge, but in opposite ways. 
Then the peak in $\varepsilon_{33,tot}$ for the nanobubble phase occurs \textit{before} the large NC magnitude increase in BTO. 
Similarly, no switching is observed at the larger compressive strain $-2.13\%$ in Figure \ref{fig:PhaseDiagram}. 
Notably, the $-2.13\%$ strain, the boundary between the nanobubble and monodomain phases, harbors the largest external dielectric permittivity amongst all strains investigated.
Reference~\cite{Kasamatsu2016} observed a similar maximal response at the multidomain-monodomain transition under an appropriate bias field in a BTO/STO superlattice, providing another illustration of how, through local field effects, misfit strain can mimic the application of an electric field~\cite{Lai2006}.

On the other hand, the NC effect in the \textit{monodomain} phase $i^{\prime}$ has a distinct mechanism. In this phase, no domain wall can move and overscreen the polarization as discussed in Refs.~\cite{Bratkovsky2001,Lukyanchuk2018,Iniguez2019}. Rather, we are working in the ``incipient ferroelectric" regime~\cite{Iniguez2019}, for which $1/\varepsilon_{33}^{int}$ should be minimum at the transition temperature $T_{\uparrow \uparrow}$ at which a monodomain polar state forms~\cite{Zubko2016}. This is approximately verified for the large compressive strain $-2.13\%$ in Figure \ref{fig:Epsilons}, where the peak of the dielectric permittivity of the BTO/STO SL coincides with the minimum in the internal dielectric permittivity of the BTO layer. We also note that, in Figure \ref{fig:Epsilons}, the NC of the BTO layer is preserved well above the nominal transition point $T_0$ of strained bulk BTO (characterized by the maximum of the dielectric permittivity in the red dashed dotted curves of the upper panels, and by the solid red line in the phase diagram in Figure \ref{fig:PhaseDiagram}), while Ref.~\cite{Zubko2016,Iniguez2019} mention its appearance only below $T_0$. As a matter of fact, within our investigated range of temperature, only tensile strains show a positive capacitance of the BTO layer at high temperature (not shown here). It could be a manifestation of the partially order-disorder character~\cite{Zalar2003,Stern2004,Hlinka2008,Ponomareva2008} 
(\textit{i.e.}, not totally displacive as considered by the models in Refs.~\cite{Zubko2016,Iniguez2019}) of the ferroelectric transition in BTO, and highlights that negative internal dielectric permittivities are linked with the relative strength of dipolar fluctuations between dielectric layers rather than a particular dipole ordering.

As for enhancing overall capacitance by NC, we return to the top panels of Figure \ref{fig:Epsilons}. At the nanobubble-monodomain transition ($-2.13\%$ strain), a significant enhancement of overall capacitance is realized relative to bulk BTO and BST, and at least eight times enhancement over bulk STO at its maximal response, compared with three times enhancement in SrTiO$_3$/PbTiO$_3$ superlattices~\cite{yadav2019spatially}.
Even where one or more of bulk BTO, STO, or BST has a greater maximum of overall dielectric response, the diffusive character of the multidomain transition in the superlattice allows for large capacitance over a much broader temperature range closer to room temperature, intermediated by a nearly constant negative dielectric response in the BTO layer over that range. 
\section{V. Conclusion}
In summary, our first-principles-based effective Hamiltonian calculations reveal the existence of negative internal dielectric permittivities in the BTO layer of a BTO/STO superlattice. These quantities are associated with (static) negative capacitance and allow the tuning of overall capacitance, which is found to be largest at the phase boundary between the ferroelectric monodomain and nanobubble states, in accordance with the optimization of other physical properties (such as piezoelectricity) along the same boundary in other nanostructures~\cite{Zhang2017}.
In addition, we predict a previously unreported switching that exchanges the negative capacitance between the BTO and STO layers at moderately low ($\sim 200$~K) temperatures and strains (approximately between $-1\%$ and $+ 1$\%), associated with both ferroelectricity of STO under compressive strain and the diffusive character of the multidomain phase transition extending to the tensile regime. 
Such a result is also true in a  (BaTiO$_3$)$_{10}$/(SrTiO$_3$)$_{10}$ superlattice with higher Sr content (not shown).
Moreover, significant enhancement of the overall capacitance is realized relative to bulk BTO, STO, and BST at the nanobubble-monodomain transition. Furthermore, relative to these bulk materials, the superlattice significantly broadens the temperature range of large (positive) overall capacitance closer to room temperature for all strains.
We hope the demonstrated strain control improves the technological and scientific viability of tunable negative capacitance devices. 
\begin{acknowledgments}
R.W. and C.P. acknowledge ARO Grant No. W911NF-16-1-0227. S. P. and L.B. thank ONR Grant No. N00014-17-1-2818. Some computations
were made possible by MRI Grant No. 0722625 from NSF, ONR Grant No. N00014-15-1-2881 (DURIP), and a Challenge grant from the Department of Defense.
S. P. appreciates support of Russian Ministry of
Science and Education (RMES) 3.1649.2017/4.6 and
RFBR 18-52-00029$\_$Bel$\_$a.
\end{acknowledgments}

$^*$ rwalter@email.uark.edu


\begin{thebibliography}{3}

\bibitem{Khan2015} A. I. Khan, K. Chatterjee, B. Wang, S. Drapcho, L. You, C. Serrao, S. R. Bakaul, R. Ramesh, S. Salahuddin, \href{http://www.nature.com/articles/nmat4148}{\textit{Nat. Mater.} \textbf{14}, 182} (2015).

\bibitem{Hoffmann2016} M. Hoffmann, M. Pe\^{s}i\'{c}, K. Chatterjee, A. I. Khan, S. Salahuddin, S. Slesazeck, U. Schroeder, T. Mikolajick, \href{http://doi.wiley.com/10.1002/adfm.201602869}{\textit{Adv. Func. Mater.} \textbf{26}, 8643} (2016).

\bibitem{Hoffmann2018} M. Hoffmann, A. I. Khan, C. Serrao, Z. Lu, S. Salahuddin, M. Pešić, S. Slesazeck, U. Schroeder, T. Mikolajick, \href{http://aip.scitation.org/doi/10.1063/1.5030072}{\textit{J. Appl. Phys.} \textbf{123}, 184101} (2018).

\bibitem{Zubko2016} P. Zubko, J. C. Wojdel, M. Hadjimichael, S. Fernandez-Pena, A. Sen\'{e}, I. Luk'yanchuk, J.-M. Triscone, J. \'{I}\~{n}iguez, \href{http://www.nature.com/articles/nature17659}{\textit{Nature} \textbf{534}, 524} (2016).

\bibitem{Lukyanchuk2019} I. Luk'yanchuk, Y. Tikhonov, A. Sen\'{e},, A. Razumnaya, V. M. Vinokur, \href{https://doi.org/10.1038/s42005-019-0121-0}{\textit{Communications Physics} \textbf{2}, 1} (2019). 

\bibitem{Salahuddin2008} S. Salahuddin, S. Datta, \href{https://pubs.acs.org/doi/10.1021/nl071804g}{\textit{Nano Lett.} \textbf{8}, 405} (2008).

\bibitem{Iniguez2019} J. \'{I}\~{n}iguez, P. Zubko, I. Luk'yanchuk, A. Cano, \href{https://doi.org/10.1038/s42005-019-0121-0}{\textit{Nat. Rev. Mater.}\textbf{4}, 1} (2019).

\bibitem{Krowne2011} C. M. Krowne, S. W. Kirchoefer, W. Chang, J. M. Pond, L. M. B. Alldredge, \href{https://pubs.acs.org/doi/10.1021/nl1037215}{\textit{Nano Lett.} \textbf{11}, 988} (2011).

\bibitem{Bratkovsky2001} A. M. Bratkovsky, A. P. Levanyuk, \href{https://link.aps.org/doi/10.1103/PhysRevB.63.132103}{\textit{Phys. Rev. B} \textbf{63}, 132103} (2001).

\bibitem{Lukyanchuk2018} I. Luk'yanchuk, A. Sen\'{e}, V. M. Vinokur, \href{https://link.aps.org/doi/10.1103/PhysRevB.98.024107}{\textit{Phys. Rev. B} \textbf{98}, 024107} (2018).

\bibitem{Ponomareva2007a} I. Ponomareva, L. Bellaiche, R. Resta, \href{https://link.aps.org/doi/10.1103/PhysRevB.76.235403}{\textit{Phys. Rev. B} \textbf{76}, 235403} (2007).

\bibitem{Ponomareva2007b} I. Ponomareva, L. Bellaiche, R. Resta, \href{https://link.aps.org/doi/10.1103/PhysRevLett.99.227601}{\textit{Phys. Rev. Lett.} \textbf{99}, 227601} (2007).

\bibitem{Walizer2006} L. Walizer, S. Lisenkov, L. Bellaiche, \href{https://link.aps.org/doi/10.1103/PhysRevB.73.144105}{\textit{Phys. Rev. B} \textbf{73}, 144105} (2006).

\bibitem{Dawber2005} M. Dawber, K. M. Rabe, J. F. Scott, \href{https://link.aps.org/doi/10.1103/RevModPhys.77.1083}{\textit{Rev. Mod. Phys.} \textbf{77}, 1083} (2005).

\bibitem{Okatan2009} M. B. Okatan, A. L. Roytburd, J. V. Mantesc, S. P. Alpay, \href{https://aip.scitation.org/doi/10.1063/1.3142385}{\textit{J. Appl. Phys.} \textbf{105}, 114106} (2009).

\bibitem{Cole2003} M. W. Cole, W. D. Nothwang, C. Hubbard, E. Ngo, M. Ervin, \href{https://aip.scitation.org/doi/10.1063/1.1569392}{\textit{J. Appl. Phys.} \textbf{93}, 9218} (2003).

\bibitem{Lisenkov2007} S. Lisenkov, L. Bellaiche, \href{https://link.aps.org/doi/10.1103/PhysRevB.76.020102}{\textit{Phys. Rev. B} \textbf{76}, 020102(R)} (2007).

\bibitem{Choudhury2011} N. Choudhury, L. Walizer, S. Lisenkov, L. Bellaiche, \href{http://www.nature.com/articles/nature09752}{\textit{Nature} \textbf{470}, 7335} (2011).

\bibitem{Lai2006}
B.K. Lai, I. Ponomareva, I.I. Naumov, I. Kornev, H. Fu, L. Bellaiche, G.J. Salamo,
\href{https://journals.aps.org/prl/pdf/10.1103/10.1103/PhysRevLett.96.137602}
{\textit{Phys. Rev. Lett.} \textbf{96} 137602} (2006).

\bibitem{Zhang2017}
Q. Zhang, L. Xie, G. Liu, S. Prokhorenko, Y. Nahas, X. Pan, L. Bellaiche, A. Gruverman and N. Valanoor, \href{https://doi.org/10.1002/adma.201702375}{\textit{Adv. Mater.} \textbf{29} 46} (2017).

\bibitem{Ponomareva2005} I. Ponomareva, I.I. Naumov, I. Kornev, H. Fu, L. Bellaiche, \href{https://link.aps.org/doi/10.1103/PhysRevB.72.140102}{\textit{Phys. Rev. B} \textbf{72}, 140102(R)} (2005).

\bibitem{frenkel2001understanding} D. Frenkel, B. Smit, \textit{Understanding molecular simulation: from algorithms to applications}, Elsevier, Amsterdam, Netherlands (2001).

\bibitem{Pertsev2000} N.A. Pertsev, A.K. Tagantsev, N. Setter, \href{https://link.aps.org/doi/10.1103/PhysRevB.61.R825}{\textit{Phys. Rev. B} \textbf{61}, R825} (2000). 

\bibitem{Davis2006} M. Davis, D. Damjanovic, N. Setter, \href{https://link.aps.org/doi/10.1103/PhysRevB.73.014115}{\textit{Phys. Rev. B} \textbf{73}, 014115} (2006). 

\bibitem{Cheng2019} P.H. Cheng, Y.T. Yin, I.N. Tsi, C.H. Lu, S.C. Pan, J. Shieh, M. Shiojiri, M.J. Chen, \href{https://doi.org/10.1038/s42005-019-0120-1}{\textit{Communications Physics} \textbf{2}, 1} (2019). 

\bibitem{Kasamatsu2016}
S. Kasamatsu, S. Watanabe, C.S. Hwang, S. Han, \href{https://doi.org/10.1002/adma.201502916}{\textit{Adv. Mater.} \textbf{28} 2} (2016).

\bibitem{Stern2004} E. A. Stern, \href{https://journals.aps.org/prl/pdf/10.1103/PhysRevLett.93.037601}{\textit{Phys. Rev. Lett.} \textbf{93}, 037601} (2004).

\bibitem{Zalar2003} B. Zalar, V. V. Laguta, R. Blinc, \href{https://journals.aps.org/prl/pdf/10.1103/PhysRevLett.90.037601}{\textit{Phys. Rev. Lett.} \textbf{90}, 037601} (2003).

\bibitem{Hlinka2008} J. Hlinka, T. Ostapchuk, D. Nuzhnyy, J. Petzelt, P. Kuzel, C. Kadlec, P. Vanek, I. Ponomareva and L. Bellaiche, \href{https://journals.aps.org/prl/pdf/10.1103/PhysRevLett.101.167402}{\textit{Phys. Rev. Lett.} \textbf{101}, 167402} (2008).

\bibitem{Ponomareva2008} I. Ponomareva, L. Bellaiche, T. Ostapchuk, J. Hlinka and J. Petzelt, \href{https://link.aps.org/doi/10.1103/PhysRevB.77.012102}{\textit{Phys. Rev. B} \textbf{77}, 012102} (2008).

\bibitem{yadav2019spatially}
A.K. Yadav, K.X. Nguyen, Z. Hong, P. Garc�a-Fern�ndez, P. Aguado-Puente, C.T. Nelson, S. Das, B. Prasad, D. Kwon, S. Cheema, A. Khan, C. Hu, J. \'{I}\~{n}iguez, J. Junquera, L.-Q. Chen, D. Muller, R. Ramesh, S. Salahuddin, \href{http://www.nature.com/articles/nature09752}{\textit{Nature} \textbf{565}, 468} (2019).

\end{thebibliography}
\end{document}